# SPATIALLY-FLAT ROBERTSON-WALKER MODELS WITH COMBINED $\Lambda CDM$ AND STIFF MATTER SOURCES AND THE CORRESPONDING THERMODYNAMICS

MARINA-AURA DARIESCU[1], DENISA-ANDREEA MIHU[1], CIPRIAN DARIESCU[1]

[1]*Faculty of Physics, "Alexandru Ioan Cuza" University of Iasi, Bd. Carol I, no. 11, 700506 Iasi, Romania, Email: marina@uaic.ro*

Abstract. In the present paper, we are considering a spatially-flat Friedmann-Robertson-Walker cosmological model, fueled with stiff matter and dust, treated as non-interacting ideal fluid sources. By solving the corresponding Friedmann equation with a non-zero cosmological constant, we are deriving the scale function and the fundamental cosmological parameters. Within a thermodynamic approach, the general form of the Equation of State is obtained, together with the explicit dependence of the energy density and pressure on temperature.

Short title: Robertson-Walker models with $\Lambda CDM$ and stiff matter sources

## 1. INTRODUCTION

Cosmological models with a stiff matter phase have always attracted considerable attention, starting with the pioneering work of Zel'dovich [1], up to recent models where a stiff fluid-like behavior is characterizing the self-interaction between dark matter components [2, 3].

After the remarkable observational discovery of the Universe acceleration was confirmed by Reiss and Perlmutter [4, 5], an active and modern branch in theoretical cosmology has focused on different matter sources, including exotic ones, that yield accelerating solutions to Einstein's equations [6]. In this respect, it has been speculated that this unanimously recognized accelerated expansion is driving by an elusive form of matter which was called dark energy. Its elucidation is expected to provide a hint on an old mystery: the gravitational effect of the zero-point energies of particles and fields [7].

As a theoretical candidate for dark energy, the cosmological constant $\Lambda$ is undoubtedly the simplest and the most appealing one. Even though, in classical Friedmann-Robertson-Walker (FRW) models with zero curvature, the accelerated phase of expansion is generally explained by this concept, the so-called cosmological constant problem remains a historical unsolved issue [8]. However, the $\Lambda$ relevance in the Universe dynamics cannot be denied, the quantum fluctuations, responsible for generating a substantial vacuum energy density, being found like acting as a cosmological constant [8]. Moreover, the inclusion of a cosmological constant term in the standard model of the Big Bang theory seems to make it consistent with the observations of large scale distribution of galaxies and clusters, with WMAP's measurements of CMB fluctuations and also with the detected properties of X-ray clusters [9].

In what it concerns alternative scenarios to reveal the nature of dark energy, we just point out the quintessence model ($w > -1$), the phantom ($w < -1$) and the quintom that, possessing a time dependent Equation of State (EoS) parameter, can pass from the phantom region to the quintessence one. Other models can be found in [10, 11].

Concerning the basic characteristics of the Universe, observations on the large scale structure which indicate its isotropy and spatial homogeneity seem to favor the FRW cosmological models. In addition, a powerful experimental evidence in terms of the detection of temperature variations in the CMB is suggesting that the Universe left the radiation era with a spatially-flat FRW geometry [12].

Based on the above arguments, in the present study, we are considering a $k = 0$ – FRW cosmological model with stiff matter and dust, treated as non-interacting fluid sources, in the presence of a non-zero cosmological constant. In what it concerns the radiation component, due to its negligible contribution to the actual energy density of the Universe, we are neglecting its presence, but we point out that its contribution must be taken into consideration in models where one is mainly investigating the early Universe stages.

The motivation for our matter choice is sustained by the intensive studies dedicated to cosmology dynamics in the presence of a stiff fluid [13, 14]. The large variety of particle species produced after the Big Bang event due to the expansion and the temperature drop of the Universe [15, 16] represents a powerful reason for which the stiff matter is considered in FRW cosmological models, in agreement to the pioneering works of Zel'dovich [1, 17]. It worth mentioning that, few years later, Zel'dovich tried to unify the structure and the entropy of the Universe based on the idea that, in case of a stiff fluid, the speed of light is equal to the speed of sound and the equations governing its dynamics can be put into correspondence with those of the gravitational field [6].

More recently, it has been suggested that the Universe passed through an early phase which was prior to the radiation dominated era, a phase which would have contained a matter composition in the form of stiff matter [18].

As for the dust component inclusion, we would like to point out its outstanding role, from the Einstein-de Sitter model, to structures formation. This type of matter has been present starting with very early phases of the universe, some galaxies being identified as containing dust particles which could lead to the formation of planets [19]. In the same time, quantities of dust were detected in the high-redshift galaxies and quasars [20] and various analytical and numerical models [21] are trying to understand its evolution in galaxies. Other main dust sources in the actual universe have been identified in terms of AGB stars and various observations come to support this idea [22]. Thus, at the theoretical level, dust evolution and its origin are still puzzling. The whole model, cosmological constant + cold dark matter as cosmologic dust, which goes by the name of $\Lambda CDM$, has been remarkably successful in describing the real Universe, from the matter-dominated phase to the recent accelerated expansion.

3Our paper is organized as it follows. Firstly, the Friedmann equation for the model proves to be analytically integrable leading to the expression of the scale factor, in agreement with the investigations of Chavanis [23]. Alternatively, one may use the GRTensorII package to study simple cosmological models [24].

The derived scale function allows us to evaluate the fundamental cosmological parameters and their behavior, both in the early and in the late Universe eras.

In the next section, we are dedicating ourselves into finding the general EoS describing the model, starting from the first law of thermodynamics. The dependence of the energy density and pressure with respect to the temperature is derived. Within a level of approximation, our EoS becomes the one of Zel'dovich's cosmological model [1, 6].

## 2. THE $k=0$ - FRW UNIVERSE WITH MIXED MATTER SOURCES

Let us consider the four-dimensional $k=0$ - FRW metric written as

$$ds^2 = a^2(t)\left[dr^2 + r^2(d\theta)^2 + r^2\sin^2\theta(d\varphi)^2\right] - dt^2 , \qquad (1)$$

where $a(t) = a_0 e^{f(t)}$ is the scale factor and $t$ is the cosmic time. For the pseudo-orthonormal frame $e_{a(a=\overline{1,4})}$, with the dual base

$$\omega^1 = a\,dr,\ \omega^2 = ar\,d\theta,\ \omega^3 = ar\sin\theta\,d\varphi,\ \omega^4 = dt,$$

the Cartan equations

$$d\omega^a = \Gamma^a_{\cdot[bc]}\omega^b \wedge \omega^c,\ \text{with}\ 1 \le b < c \le 4;$$
$$R_{ab} = d\Gamma_{ab} + \Gamma_{ac} \wedge \Gamma^c_{\cdot b} \qquad (2)$$

lead to the connection one-forms $\Gamma_{ab} = \Gamma_{abc}\omega^c$, i.e.

$$\Gamma_{14} = \frac{\dot{a}}{a}\omega^1,\ \Gamma_{24} = \frac{\dot{a}}{a}\omega^2,\ \Gamma_{34} = \frac{\dot{a}}{a}\omega^3,$$
$$\Gamma_{21} = \frac{1}{ar}\omega^2,\ \Gamma_{31} = \frac{1}{ar}\omega^3,\ \Gamma_{32} = \frac{\cot\theta}{ar}\omega^3,$$

and to the curvature two-forms

$$R_{ab} = \frac{1}{2}R_{abcd}\omega^c \wedge \omega^d .$$

With the derived Ricci tensor components, $R_{ab} = R^c_{\cdot acb}$, i.e.

$$R_{\alpha\beta} = \left[\frac{\ddot{a}}{a} + 2\left(\frac{\dot{a}}{a}\right)^2\right]\delta_{\alpha\beta}; \quad R_{44} = -3\frac{\ddot{a}}{a}, \tag{3}$$

where $\alpha = 1, 2, 3$ and the Ricci scalar

$$R = 6\left(\frac{\dot{a}^2}{a^2} + \frac{\ddot{a}}{a}\right), \tag{4}$$

the Einstein tensor components read

$$G_{\alpha\alpha} = -2\frac{\ddot{a}}{a} - \left(\frac{\dot{a}}{a}\right)^2 \; ; \quad G_{44} = 3\frac{\dot{a}^2}{a^2}, \tag{5}$$

where the overdot means the derivative with respect to the cosmic time $t$.

As for the matter source, let us consider a perfect fluid of energy-momentum tensor $T_{\alpha\alpha} = p$, $T_{44} = \rho$, so that the Einstein's system of equations,

$$G_{ab} + \eta_{ab}\Lambda = \kappa T_{ab}, \tag{6}$$

gets the explicit form

$$2\frac{\ddot{a}}{a} + \frac{\dot{a}^2}{a^2} - \Lambda = -\kappa p;$$
$$3\frac{\dot{a}^2}{a^2} - \Lambda = \kappa \rho \tag{7}$$

where $\kappa = 8\pi G / c^4$ and $\Lambda$ is the positive cosmological constant.

In the followings, we are going to focus on the second relation in (7), known as the Friedmann's equation for $k = 0$ - FRW Universe,

$$H^2 = \frac{\kappa}{3}\rho + \frac{\Lambda}{3}, \tag{8}$$

where $H$ is the Hubble function defined as $H = \dot{a}/a$.

Following a similar analytical endeavor as in the paper of Chavanis [25], we are considering a cosmological model of Universe whose composition is given by mixed matter non-miscible sources: stiff matter and dust.

Let us remind the reader that the so-called stiff matter era has been proposed more than 50 years ago, in Zel'dovich' studies [1, 6], where the author developed a cosmological model for a very early stage of the Universe filled with a cold gas of baryons interacting through a meson field. The derived EoS was describing the four phases of the Universe (stiff matter phase, radiation phase, dust matter era and dark energy phase) simultaneously present. Recently, a stiff matter epoch has been considered in cosmological models with dark matter made of self-gravitating Bose–Einstein condensates (BECs) [25].

For the Universe filled with a combined matter source made of non-interacting stiff matter and dust, with the total energy density given by

$$\rho = \rho_{01}\left(\frac{a_0}{a}\right)^6 + \rho_{02}\left(\frac{a_0}{a}\right)^3 = \frac{3}{\kappa}\left(\frac{\alpha}{a^6} + \frac{\beta}{a^3}\right), \qquad (9)$$

the general Friedmann equation (8) takes the explicit form

$$\dot{a}^2 = \frac{\alpha}{a^4} + \frac{\beta}{a} + a^2\lambda. \qquad (10)$$

We have introduced the notations

$$\alpha = \frac{\kappa}{3}\rho_{01}a_0^6, \ \beta = \frac{\kappa}{3}\rho_{02}a_0^3, \ \lambda = \frac{\Lambda}{3}, \qquad (11)$$

where the zero index corresponds to the present day values.

The evolution of the scale factor is obtained by integrating the equation (10), its closed form analytic solution being

$$a(t) = \left[\frac{\beta}{2\lambda}\cosh(3\sqrt{\lambda}t) + \sqrt{\frac{\alpha}{\lambda}}\sinh(3\sqrt{\lambda}t) - \frac{\beta}{2\lambda}\right]^{1/3}, \qquad (12)$$

where the integration constant has been found from the physical constraint in the origin of the scale factor, $a(0) = 0$. For $t \to 0$, by expanding in power series the hyperbolic functions, one gets the asymptotic expression of the scale function

$$a_{t\to 0} \approx \left(3\sqrt{\alpha}t\right)^{1/3}\left[1 + \frac{\beta}{4\sqrt{\alpha}}t + O(t^2)\right], \qquad (13)$$

while for large values of the argument ($t \to \infty$), one recovers the de Sitter solution $a_{t\to\infty} \sim e^{\sqrt{\lambda}t}$, in agreement with the expansion concept and, particularly, with the current validated idea of a continuously expanding Universe [4, 5]. This is also confirmed by the positive values of the Hubble parameter computed with (12) as

$$H(t) = \lambda^{1/2}\frac{\beta\sinh\left(3\sqrt{\lambda}t\right) + 2\sqrt{\alpha\lambda}\cosh\left(3\sqrt{\lambda}t\right)}{\beta\left[\cosh\left(3\sqrt{\lambda}t\right) - 1\right] + 2\sqrt{\alpha\lambda}\sinh\left(3\sqrt{\lambda}t\right)}. \qquad (14)$$

Concerning the asymptotic representations of the expression in (14), a simple algebraic computation provides us with $H \sim 1/(3t)$, for small $t$ values and with $H \sim \lambda^{1/2} = const.$, for large values of the time variable.

Recalling again the Friedmann equation (8), we are able to determine the energy density

$$\rho(t) = \frac{3}{\kappa}\lambda\left\{\left[\frac{\beta\sinh(3\sqrt{\lambda}t) + 2\sqrt{\alpha\lambda}\cosh(3\sqrt{\lambda}t)}{\beta\left[\cosh(3\sqrt{\lambda}t) - 1\right] + 2\sqrt{\alpha\lambda}\sinh(3\sqrt{\lambda}t)}\right]^2 - 1\right\}, \quad (15)$$

which, near the time origin blows up as

$$\rho_{t\to 0} \approx \frac{1}{\kappa}\left[\frac{1}{3t^2} + \frac{\beta}{2\sqrt{\alpha}t} - \frac{3\beta^2}{16\alpha} - \lambda + O(t)\right], \quad (16)$$

while, in the far future, goes exponentially to zero. As for the pressure, this can be derived from the first Einstein's equation in (7) as depending on the scale factor as

$$p(a) = \frac{3\alpha}{\kappa a^6}, \quad (17)$$

pointing out a stiff matter behavior. One may notice that the time dependence of the expression (17), i.e.

$$p(t) = \frac{12\alpha}{\kappa}\left[\frac{\lambda}{\beta\cosh(3\sqrt{\lambda}t) + 2\sqrt{\alpha\lambda}\sinh(3\sqrt{\lambda}t) - \beta}\right]^2, \quad (18)$$

goes exponentially to zero, for $t \to \infty$, and can be approximated to

$$p_{t\to 0} \approx \frac{1}{\kappa}\left[\frac{1}{3t^2} - \frac{\beta}{2\sqrt{\alpha}t} + \frac{9\beta^2}{16\alpha} - \lambda + O(t)\right], \quad (19)$$

for small $t$ values.

A basic analysis in cosmology involves the study of the dynamics of the EoS parameter relating the pressure and the energy density by the well-known relation $p = w\rho$. In the case under consideration, for $t \to 0$, with invoking the series expansions (16) and (19), one is able to deduce, in the first order of approximation, the following linear time-dependence

$$w(t) = \frac{p_{t\to 0}}{\rho_{t\to 0}} \approx 1 - \frac{3\beta}{\sqrt{\alpha}}t + O(t^2), \quad (20)$$

from which the single type of matter Universe eras are to be recovered. In this respect, one may notice that $w$ is inferior bounded by $w=0$, corresponding to a dust matter domination phase, at the instant $t=\frac{\sqrt{\alpha}}{3\beta}$ and superior bounded by $w=1$, at the $t=0$, characterizing a stiff matter era. Between these two pivot values of time, $w$ takes all the positive values in the range $w \in [0,1]$, as for example $w=1/3$, at $t=\frac{2\sqrt{\alpha}}{9\beta}$, characterizing the radiation era. This result is sustained by the large number of studies reporting a time-dependent EoS, as for example the models with viscous fluids [26] or quintessence models with scalar fields [27, 28]. We remark that most of these studies are involving not only positive $w$ values, but also negative ones. As for the correlation with observations, we recommend [29].

At the end of this section, let us discuss the present age of the Universe characterized by the scale function (12). In literature, this is computed as the following integral [23]

$$\tau = \int_0^\tau dt = \frac{1}{H_0} \int_0^1 \frac{du}{u\sqrt{\frac{\Omega_1^0}{u^6}+\frac{\Omega_2^0}{u^3}+\Omega_\Lambda}}, \qquad (21)$$

where $u=a/a_0$ and the above parameters are correlated by the flatness property

$$\Omega_1^0 + \Omega_2^0 + \Omega_\Lambda = 1. \qquad (22)$$

This is consistent with recent CMB measurements of MAXIMA-I flight and COBE-DMR experiment [30], sustaining the idea that our Universe is flat, in accordance with most of the inflationary models [31].

Coming back to our model, the present age of the Universe can be estimated from the relation

$$\tau = \int_0^\tau dt = \frac{1}{H_0} \int_0^1 \frac{du}{u\sqrt{\frac{\alpha}{u^6 a_0^6}+\frac{\beta}{u^3 a_0^3}+\lambda}}, \qquad (23)$$

and one may identify the parameters defined in (11) as being related to the values of $\Omega_1^0$, $\Omega_2^0$ and $\Omega_\Lambda$, by

$$\frac{\alpha}{a_0^6}=\Omega_1^0, \ \frac{\beta}{a_0^3}=\Omega_2^0, \ \lambda=\Omega_\Lambda, \qquad (24)$$

with the constraint (22). The numerical values for the dimensionless densities $\Omega_1^0$, $\Omega_2^0$, $\Omega_\Lambda$ are determined experimentally from measurements of type Ia supernovas (SN Ia's) [32]. These observations, correlated with gravitational lenses and stellar dynamical analysis, report the numerical values $\Omega_{matter} \sim 0.237$ and $\Omega_\Lambda \sim 0.763$ [33]. As assumed in [23], $a_0 = 1.32 \cdot 10^{26} m$ and the matter component value is distributed as $\Omega_1^0 \sim 0.001$ for the stiff component and $\Omega_2^0 \sim 0.237$ for the dust one.

Putting everything together, we find, by integration the present age of the Universe as being given by the relation

$$\tau = \frac{1}{3H_0} \frac{1}{\sqrt{1-\Omega_1^0-\Omega_2^0}} \ln\left[\frac{2(1-\Omega_1^0)-\Omega_2^0+2\sqrt{1-\Omega_1^0-\Omega_2^0}}{\Omega_2^0+2\sqrt{\Omega_1^0(1-\Omega_1^0-\Omega_2^0)}}\right]. \qquad (25)$$

By using in the relation (25), the most up-to-date measurements data for the current age of the Universe, $\tau = 13,79 \times 10^9 \, yrs$, and for the Hubble parameter, $H_0 = 67,8 \, km \cdot s^{-1} \cdot Mpc^{-1}$, reported in [34], we get the following transcendental relation between the dimensionless parameters $\Omega_1^0$ and $\Omega_2^0$, which are constraining the values of the model's parameters $\alpha$ and $\beta$,

$$\frac{2(1-\Omega_1^0)-\Omega_2^0+2\sqrt{1-\Omega_1^0-\Omega_2^0}}{\Omega_2^0+2\sqrt{\Omega_1^0(1-\Omega_1^0-\Omega_2^0)}} = \exp\left(2.86\sqrt{1-\Omega_1^0-\Omega_2^0}\right). \qquad (26)$$

Thus, one may define the function

$$f(x,y) = \frac{2(1-x)-y+2\sqrt{1-x-y}}{y+2\sqrt{x(1-x-y)}} \exp\left(-2.86\sqrt{1-x-y}\right) \qquad (27)$$

which is represented below, for allowed ranges of $x = \Omega_1^0$ and $y = \Omega_2^0$. The figure 1 offers us the allowed pairs $(\Omega_1^0, \Omega_2^0)$ given by the curve drawn by the intersection of the two-dimensional sheet representing the function (27) and the horizontal plane $f(\Omega_1^0, \Omega_2^0) = 1$, imposed by the data. For example, to $x = \Omega_1^0 = 0.001$, the corresponding value is $y = \Omega_2^0 = 0.232$.

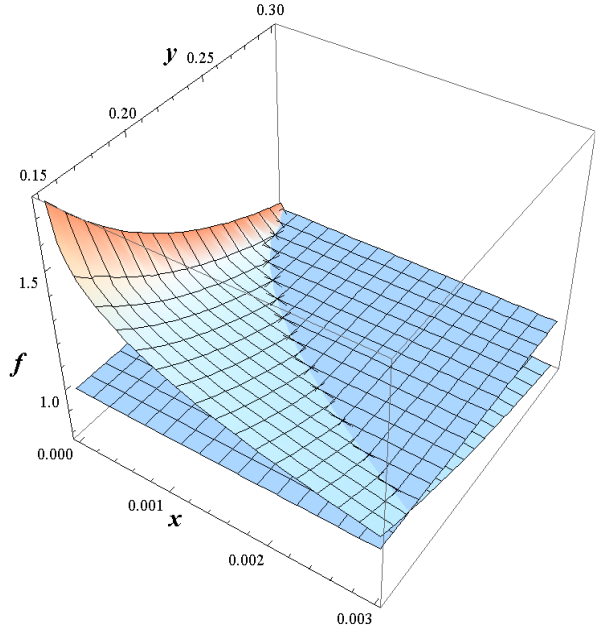

**Figure 1** - The 3D plot of the function $f(x,y)$ given in (27) and its intersection with the unity plane $f(x,y)=1$.

## 3. GENERAL EQUATION OF STATE AND ENTROPY

The form (17) of the pressure together with the density energy (9) lead to the relation

$$\rho(p) = p + \beta\sqrt{\frac{3p}{\kappa\alpha}} \qquad (28)$$

whose inversion allows us to find the proper EoS governing our cosmological model, namely the interesting negative branch (of the algebraic solution):

$$p(\rho) = \rho - \beta\sqrt{\frac{3}{\kappa\alpha}}\sqrt{\rho + \frac{3\beta^2}{4\kappa\alpha}} + \frac{3\beta^2}{2\kappa\alpha}. \qquad (29)$$

One may notice in (29), besides the linear contribution corresponding to a stiff matter source, $p \sim \rho$, the polytropic square-root term $p \sim \sqrt{\rho}$, which is similar to the one describing the relativistic Bose-Einstein condensates (BEC) model developed in [25]. One may find a detailed treatment of the cosmic BEC mechanism in [35].

By performing a series expansion of the square root in the equation (29), one can write down a more physically interpretable formulation for the derived EoS, i.e.

$$p(\rho) \approx \frac{\kappa\alpha}{3\beta^2}\rho^2 + O(\rho^3). \qquad (30)$$

This result agrees with the quadratic expression $p(\rho) = K\rho^2$ derived by Zel'dovich who has attached an additional term responsible for quantum corrections to describe a model where the four phases of the Universe are simultaneously present. Also, the EoS (30) can be put into correspondence with the one for the BEC derived from Gross-Pitaevskii (GP) equation [36]

$$p(\rho) = \frac{2\pi\hbar^2 l_s}{m^3}\rho^2, \qquad (31)$$

where $\hbar$ is the reduced Planck constant, $l_s$ represents the scattering length, $m$ defines the mass of particles subjected to the condensation process and $\rho$ is the density distribution characterizing one single component of dark matter BEC. The generalized form of (31), i.e. $p(\rho) = w\rho + k\rho^2$, does also contain a linear term and has been recently used to describe the dynamics of the very early phase of the Universe [37].

To conclude, the relation (29) turns, for $\rho \to 0$ (the non-relativistic limit), into the quadratic form, $p \sim \rho^2$, compatible with the polytropic EoS for BEC stars. In the early stages (the ultra-relativistic limit), we are left with the stiff matter contribution, $p \sim \rho$.

Moreover, the equation (29) allows an interpretation within the frame of cosmological viscous effects. Recently, it has been suggested that the background dynamics of a homogeneous and isotropic Universe can be driven by a cosmological bulk viscosity which contributes to the cosmic fluid pressure [38]. More exactly, within the cosmological evolution scenario, the bulk viscosity is compatible with an effective pressure, its role being to recover the thermal equilibrium state of the system, which has been destroyed during the very fast expansion of the Universe. Thus, for the following relation between the model parameters

$$3\beta^2 = 4\alpha\Lambda, \qquad (32)$$

the EoS (29) becomes

$$p = \rho - \frac{6}{\kappa}\sqrt{\frac{\Lambda}{3}}H + \frac{2\Lambda}{\kappa}, \qquad (33)$$

its form being similar to the one of a viscous fluid $p = \rho - 3H\zeta$, with the bulk viscosity given by

$$\zeta = \frac{2}{\kappa}\sqrt{\frac{\Lambda}{3}} = \frac{\beta}{\kappa\sqrt{\alpha}}.$$

The expression (32), in view of (24), leads to the following equation characterizing the distribution of matter between the stiff component and the dust one

$$4\Omega_1^0\left(1-\Omega_1^0-\Omega_2^0\right)=\left(\Omega_2^0\right)^2,$$

whose physical solution is

$$\Omega_1^0 = \frac{1}{2}\left(1-\Omega_2^0-\sqrt{1-2\Omega_2^0}\right) \approx \frac{\left(\Omega_2^0\right)^2}{4}\left(1+\Omega_2^0\right). \tag{34}$$

In what it concerns the thermodynamic analysis, this can be done by employing the equation [39]

$$\frac{dp}{dT} = \frac{1}{T}(\rho+p), \tag{35}$$

provided by the first principle of thermodynamics, coupled to the relation (29) between pressure and energy density, characterizing our Universe model. Thus, we arrive at the following pressure-temperature dependence,

$$p(T) = \frac{1}{4}\left[\eta T - \beta\sqrt{\frac{3}{\kappa\alpha}}\right]^2 = \frac{3\beta^2}{4\kappa\alpha}\left[\frac{T}{T_0}-1\right]^2, \tag{36}$$

where the integration constant $\eta$ has been associated to the minimum temperature $T_0$ by

$$\eta \equiv \frac{\beta}{T_0}\sqrt{\frac{3}{\kappa\alpha}}.$$

The equation (36) is a quadratic one containing, besides the $T^2$ contribution, a linear term and a constant one. With reference to the latter, this can be regarded as an effective cosmological constant or a so-called *bag pressure constant*, introduced as the effect of the quantum-chromodynamical vacuum, in the quark-hadron (de)confinement phase transition when the quarks and gluons coagulate to form hadrons [40].

The thermodynamic relation for the energy density coming from (28) is

$$\rho(T) = \frac{3\beta^2}{4\kappa\alpha}\left(\frac{T^2}{T_0^2}-1\right) \tag{37}$$

and we notice the quadratic dependence with respect to the temperature. Finally, these results lead to the scale factor expression, in terms of temperature,

$$a(T) = \left[\frac{2\alpha}{\beta}\right]^{1/3} \left[\frac{T}{T_0} - 1\right]^{-1/3} \tag{38}$$

and to the entropy of the Universe which keeps the constant value

$$S = \frac{a^3}{T}(\rho + p) = \frac{3\beta}{\kappa\alpha T_0}, \tag{39}$$

depending on the ratio of the model's parameters $\beta$ and $\alpha$, characterizing the dust and stiff matter sources.

## 4. CONCLUSIONS

For the four-dimensional $k = 0$ - FRW metric (1), we have used the Cartan's formalism to derive the main geometrical features. For a combined matter source made of stiff matter and dust, we have analytically solved the Friedmann equation and derived the expression of the scale function (12). The model parameters are constrained by the current (experimental) value of the age of the Universe. It turns out that the scale function (12) is describing a Universe in eternal expansion, from the initial singular point, to a de Sitter behavior, in the far future, which is one of the important features of the $\Lambda CDM$ model.

In early stages, the time-depending EoS parameter takes all the numerical values between $w = 1$ and $w = 0$, including the landmark value $w = 1/3$, characterizing the radiation era.

As for the general EoS (21), this contains, besides the linear contribution corresponding to a stiff matter source, the polytropic non-linear term $p \sim \sqrt{\rho}$.

If the model parameters are obeying the relation (32), the EoS turns into the form (33) characterizing a viscous fluid.

Finally, based on the first principle of thermodynamics, we have derived the dependence of the energy density, pressure and scale function on temperature and the constant value of entropy.